\documentclass[preprint,showpacs,preprintnumbers,amsmath,amssymb,floatfix]{revtex4}

\usepackage{graphicx}
\usepackage{bm}

\begin{document}

\preprint{DESY~13--159
\hspace{12.3cm}ISSN 0418-9833}
\preprint{September 2013\hspace{14.75cm}}

\title{Inclusive $\boldsymbol{J/\psi}$ and $\boldsymbol{\psi(2S)}$ production
from $\boldsymbol{b}$-hadron decay in $\boldsymbol{p\bar{p}}$ and
$\boldsymbol{pp}$
collisions}

\author{Paolo Bolzoni}
\email{paolo.bolzoni@desy.de}
\author{Bernd A. Kniehl}
\email{kniehl@desy.de}
\author{Gustav Kramer}
\email{gustav.kramer@desy.de}
\affiliation{{II.} Institut f\"ur Theoretische Physik,
Universit\"at Hamburg, Luruper Chaussee 149, 22761 Hamburg, Germany}

\date{\today}

\begin{abstract}
We study the inclusive production of $J/\psi$ and $\psi(2S)$ mesons originating
from the decays of bottom-flavored hadrons produced in $p\bar{p}$ collisions at
the Fermilab Tevatron and in $pp$ collisions at the CERN LHC.
We work at next-to-leading order in the general-mass variable-flavor-number
scheme (GM-VFNS) implemented with nonperturbative fragmentation functions
fitted to $e^+e^-$ data of inclusive $b$-hadron production exploiting their
universality.
The three-momentum distributions of the charmonia used were extracted from
$B$-decay data in the framework of nonrelativistic-QCD factorization.
Comparing the theoretical predictions thus obtained with transverse-momentum
distributions measured by the CDF~II, ALICE, ATLAS, CMS, and LHCb
Collaborations, we find excellent overall agreement as for both absolute
normalization and lineshape, which provides a nontrivial test of the GM-VFNS
over wide ranges of center-of-mass energy, transverse momentum, and rapidity.
\end{abstract}

\maketitle

\section{Introduction}

Already several years ago, the CDF Collaboration at the Fermilab Tevatron
extracted individual cross sections for the inclusive production of $J/\psi$
and $\psi(2S)$ mesons originating from decays of $B$ mesons and other $b$
hadrons \cite{1}.
The cross sections were differential in the charmonium transverse momentum
($p_T$) and covered the range 5~GeV${}<p_T<20$~GeV.
Next-to-leading-order (NLO) predictions provided by two of us \cite{2} were
found to nicely reproduce these measurements over the whole $p_T$ range.
The calculation had two ingredients, the inclusive production cross section of
the process $p\bar{p}\to B+X$, differential in $p_T$ and rapidity ($y$), and
the partial widths of the inclusive decays $B\to J/\psi+X$ and $B\to\psi(2S)+X$
as functions of the $J/\psi$ and $\psi(2S)$ momentum fractions, respectively.
The first ingredient was calculated at NLO in the zero-mass
variable-flavor-number scheme (ZM-VFNS) \cite{4}, which corresponds to the
conventional parton-model approach endowed with nonperturbative fragmentation
function (FFs) for the transition $b\to B$, as described in Ref.~\cite{3}.
In this approach, the $b$ quark is included among the incoming partons, along
with the $u$, $d$, $s$, and $c$ quarks and the gluon $g$, leading to additional
contributions.
Previous CDF measurements of the inclusive $B^+/B^0$ production cross section
at center-of-mass energy $\sqrt{s} =1.8$~TeV \cite{5} were found to be in
satisfactory agreement with such NLO ZM-VFNS predictions, provided that
realistic FFs are adopted \cite{3}.
The second ingredient was obtained in the framework of the parton model in
combination with nonrelativistic-QCD (NRQCD) factorization \cite{Bodwin:1994jh}
by applying the approach of Palmer, Paschos, and Soldan \cite{7} to the
$B\to J/\psi+X$ and $B\to\psi(2S)+X$ decay distributions measured by the CLEO
Collaboration \cite{6}.
Subsequently, the inclusive cross section of nonprompt $J/\psi$ hadroproduction
at Tevatron energies was also computed in the FONLL and MC@NLO approaches 
\cite{7a}.

The CDF Collaboration repeated their measurement of the inclusive cross section
of nonprompt $J/\psi$ \cite{8} and $\psi(2S)$ \cite{26} hadroproduction in
run~II (CDF~II) at $\sqrt{s}=1.96$~TeV with a much higher accuracy reaching
also below $p_T=5$~GeV.
Recently, all four LHC experiments, CMS \cite{9,10}, LHCb \cite{11,27}, ATLAS
\cite{12}, and ALICE \cite{13}, released their measurements of the
corresponding $J/\psi$ \cite{9,10,11,12,13}, and $\psi(2S)$ \cite{10,27}
observables in $pp$ collisions with $\sqrt{s}=7$~TeV.
These data offer the opportunity to test the $b$-hadron production models in a
new energy regime using the common decay channels to $J/\psi$ and $\psi(2S)$
mesons.

In this paper, we present a new analysis of the inclusive cross
sections of nonprompt $J/\psi$ and $\psi(2S)$ hadroproduction with theoretical
input improved relative to our previous work~\cite{2}.
Specifically, the ZM-VFNS is replaced by the general-mass
variable-flavor-number scheme (GM-VFNS), which has been elaborated in recent
years \cite{16,14,15}.
Furthermore, we adopt an updated $b\to B$ FF extracted \cite{14} from more
recent data of $e^+e^-\to B+X$ at the $Z$-boson resonance \cite{21,22,23} as
well as state-of-the-art parton distribution functions (PDFs) \cite{18}.
On the other hand, the formalism for the description of the inclusive 
decays $B \to J/\psi+X$ and $B\to \psi(2S)+X$ is taken over from
Ref.~\cite{2} without changes, since it is still quite appropriate.
To gain confidence in the reliability of our NLO treatment of inclusive
$B$-meson production, we performed comparisons \cite{14,15} with CDF~II data
from $p\bar{p}$ collisions at $\sqrt{s}=1.96$~TeV \cite{8} and with CMS data
from $pp$ collisions at $\sqrt{s}=7$~TeV \cite{17}, to find very good
agreement, in particular for larger $p_T$ values.
In Ref.~\cite {2}, the polarization of the $J/\psi$ and $\psi(2S)$ mesons from 
$b$-hadron decay was not considered.
According to the leading-order (LO) NRQCD analysis of Ref.~\cite{17b}, it is
small in both cases, which is in line with the measurement by the CDF
Collaboration \cite{Abulencia:2007us}, but in mild contrast to the one by the
BaBar Collaboration \cite{25}.

This paper is organized as follows.
In Sec.~\ref{sec:two}, we briefly describe our theoretical framework and choice
of inputs, pointing towards the appropriate references.
In Sec.~\ref{sec:three}, we compare our NLO GM-VFNS predictions for the
inclusive cross sections of nonprompt $J/\psi$ and $\psi(2S)$ hadroproduction
with recent measurements at the Tevatron \cite{8,26} and the LHC
\cite{9,10,11,27,12,13}.
Section~\ref{sec:four} contains our conclusions.

\section{Setup and input}
\label{sec:two}

The technical details of the GM-VFNS framework and results obtained from it
were previously presented in Refs.~\cite{16,14,15}.
Here, we only describe our choice of input for the numerical analysis of
nonprompt $J/\psi$ and $\psi(2S)$ hadroproduction.
We use the set CTEQ6.6 \cite{18} of proton PDFs as implemented in the LHAPDF
library \cite{19}.
This PDF set was obtained in a general-mass scheme using the input values
$m_c=1.3$~GeV, $m_b=4.5$~GeV, and $\alpha_s(m_Z)=0.118$, and taking the
starting scale of the $b$-quark PDF to be $\mu_0=m_b$.
We employ the nonperturbative $B$-meson FFs determined in Ref.~\cite{14} by
fitting experimental data on the inclusive cross section of $B$-meson
production in $e^+e^-$ annihilation taken by the ALEPH \cite{21} and OPAL
\cite{22} Collaborations at CERN LEP1 and by the SLD Collaboration \cite{23}
at SLAC SLC.
These FFs supersede the ones extracted from OPAL data \cite{20} in
Ref.~\cite{3}.
All these data were taken on the $Z$-boson resonance, so that finite-$m_b$
effects can safely be neglected.
In Ref.~\cite{14}, the asymptotic scale parameter was taken to be
$\Lambda_{\overline{\mathrm{MS}}}^{(5)}=0.227$~GeV at NLO, the factorization
and renormalization scales were identified with the $Z$-boson mass,
$\mu_F=\mu_R=m_Z$, and the starting scale of the $b\to B$ FF was chosen to be
$\mu_0 = m_b$ in accordance with Ref.~\cite{18}, while the $q,g\to B$ FFs,
where $q=u,d,s,c$, were assumed to vanish at $\mu_F=\mu_0$.
We select the FF set implemented with the simple power ansatz, which yielded
the best fit, as may be seen in Fig.~1 of Ref.~\cite{14}.
The OPAL \cite{22} and SLD \cite{23} data included all the $b$-hadron final
states, i.e.\ all the $B$ mesons, $B^\pm$, $B^0/\bar{B}^0$, and
$B_s^0/\bar{B}_s^0$, and the $b$ baryons, such as the $\Lambda_b^0$ baryon,
while, in the ALEPH analysis \cite{21}, only final states with identified
$B^\pm$ and $B^0/\bar{B}^0$ mesons were taken into account.
In Ref.~\cite{14}, the FFs of all $b$ hadrons were assumed to have the same
shape.
In addition, we shall assume here that all the $b$ hadrons have the same
branching fractions and decay distributions into $J/\psi$ and $\psi(2S)$ mesons
as the $B$ mesons.
Differences only arise from the different $b$-quark to $b$-hadron branching
fractions, which we adopt from the Particle Data Group (PDG) \cite{24}.
For example, the $B^0/\bar{B}^0$-meson contribution is to be multiplied by 
$100\%/40.1\%=2.49$.
For simplicity, we take the initial- and final-state factorization scales,
entering the PDFs and FFs, respectively, to have the same value $\mu_F$.
We choose $\mu_F$ and the renormalization scale $\mu_R$, at which $\alpha_s$ is
evaluated, to be $\mu_F=\xi_F m_T$ and $\mu_R=\xi_R m_T$, respectively, where
$m_T=\sqrt{p_T^2+m_b^2}$ with $p_T$ being the transverse momentum of the
$J/\psi$ or $\psi(2S)$ mesons, and independently vary the parameters $\xi_F$
and $\xi_R$ about their default values $\xi_F=\xi_R=1$ up and down by a factor
of two under the restriction $1/2\le\xi_R/\xi_F\le2$ to estimate the
theoretical uncertainty due to the lack of knowledge of beyond-NLO corrections.
In fact, scale variations constitute the overwhelming source of theoretical
uncertainties in our predictions.
We may, therefore, neglect the uncertainties in the PDFs and $m_b$.
For consistency with Ref.~\cite{18}, we use $m_b=4.5$~GeV throughout this work.
As in Ref.~\cite{2}, we employ an effective FF for the transition of parton $i$
via the $B$ meson to the $J/\psi$ meson, which is calculated as the convolution
\begin{eqnarray}
D_{i\to J/\psi}(x,\mu_F)=\int^{1}_{x}\frac{dz}{z}
D_{i\to B}\left(\frac{x}{z},\mu_F\right)
\frac{1}{\Gamma_B}\,\frac{d\Gamma}{dz}(z,P_B),
\label{eq:ff}
\end{eqnarray}
where $D_{i \to B}(y,\mu_F)$ are the nonperturbative FFs at $B$-to-$i$
longitudinal-momentum fraction $y$ and factorization scale $\mu_F$, as
determined in Ref.~\cite{14}, $\Gamma_B$ is the $B$-meson total decay width,
and $d\Gamma(z,P_B)/dz$ is the $B\to J/\psi$ decay distribution differential
in the $J/\psi$-to-$B$ longitudinal-momentum fraction $z$, as given in
Eqs.~(3.12) or (3.16) of Ref.~\cite{2}.
For given $J/\psi$ transverse momentum $p_T$ and rapidity $y$, the modulus of
the $B$ three-momentum $\mathbf{P_B}$ is
$P_B=|\mathbf{P_B}|=\sqrt{p_T^2+m_T^2\sinh^2y}/z$.
We use the $B^+/B^0$ average mass value $M_B=5.279$~GeV and average lifetime
value $\tau_B= 1.61$~ps. 
In Ref.~\cite{2}, the decay distribution $d\Gamma/dk_L^\prime$ in the
component $k_L^\prime$ of the $J/\psi$ three-momentum parallel to
$\mathbf{P_B}$ is obtained by integrating the general formula, given in
Eq.~(3.4) of Ref.~\cite{2}, over the orthogonal three-momentum components.
This leads to the quantity $d\Gamma(z,P_B)/dz$ appearing in Eq.~(\ref{eq:ff}),
where $z=k_L^\prime/P_B$.
It depends on the structure function $f(x)$ of the $b\to B$ transition, the
element $V_{cb}$ of the Cabibbo-Kobayashi-Maskawa matrix, and the coefficients
$a$ and $b$, which in turn depend on the short-distance coefficients of the
weak-interaction Hamiltonian of the $b\to c\bar{c}q$ transition and the
relevant $J/\psi$ long-distance matrix elements (LDMEs) of NRQCD as specified
in Eq.~(3.2) of Ref.~\cite{2}.
In Ref.~\cite{2}, the LDMEs were fitted at LO in NRQCD to the inclusive cross
section of direct $J/\psi$ hadroproduction measured by the CDF Collaboration
\cite{1} and the $B\to J/\psi+X$ branching fraction measured by the CLEO
Collaboration \cite{6}. 
The resulting prediction for the $B\to J/\psi+X$ three-momentum distribution
was found \cite{2} to be in reasonable agreement with the CLEO measurement
\cite{6}.
The latter also nicely agrees with the BaBar measurement \cite{25}, which was
not yet available for the fit \cite{2}.
Recently, NRQCD factorization has been impressively consolidated at NLO
\cite{Butenschoen:2009zy} by a global fit \cite{Butenschoen:2011yh} to the
world data on the unpolarized $J/\psi$ yields in hadroproduction,
photoproduction, two-photon scattering, and $e^+e^-$ annihilation.  
The $J/\psi$ LDMEs of Refs.~\cite{2,Butenschoen:2011yh} agree in magnitude
typically within a factor of three and in sign.
As for the color-octet LDMEs, the LO values of Ref.~\cite{2} overshoot the
respective NLO values of Ref.~\cite{Butenschoen:2011yh}, which is in line with
the observation \cite{Butenschoen:2009zy} that the NLO corrections generally
enhance the cross section of inclusive $J/\psi$ hadroproduction. 
We conclude that an update of the NRQCD analysis of the $B\to J/\psi+X$
three-momentum distribution would essentially reproduce the result of
Ref.~\cite{2}, the more so as the modelling of this decay distribution is
almost irrelevant, at least at large values of $p_T$, where the fine details
are effectively washed out by the Lorentz boost from the $B$-meson rest frame
to the laboratory frame of the hadron collider \cite{2} and the $B\to J/\psi+X$
branching fraction becomes the key parameter.
Nevertheless, we must bear in mind that NRQCD factorization is presently
challenged at NLO \cite{Butenschoen:2011ks} by LHC and Tevatron measurements of
$J/\psi$ polarization observables.
For a very recent review, we refer to Ref.~\cite{Butenschoen:2012qr}.

Besides direct $J/\psi$ production via $B\to J/\psi+X$, we also included the
feed-down contributions from $B\to\chi_{c_J}+X$ with $J=0,1,2$ followed by
$\chi_{c_J}\to J/\psi+\gamma$ and from $B\to\psi(2S)+X$ followed by
$\psi(2S)\to J/\psi+X$.
The branching fraction of the direct channel was found to be 0.80\%, while
those of the cascades via the $\chi_{c_J}$ and $\psi(2S)$ mesons were found to
be 0.13\% and 0.19\%, respectively.
Alternative LO-NRQCD analyses of the direct $B\to J/\psi+X$ branching fraction,
based on different $J/\psi$ LDME sets, yield values in the same ball park,
namely 0.65\% \cite{7} and 0.77\% \cite{17b}.
Further details may be found in Ref.~\cite{2}.
Since the appearance of the CLEO paper \cite{6}, some of these input values
have changed slightly.
However the most relevant result, namely the total $B\to J/\psi+X$ branching
fraction, goes unchanged, if up-to-date input data from the PDG \cite{24} is
used.
To facilitate the calculation, we evaluate $d\Gamma(z,P_B)/dz$ using its
asymptotic expression, obtained from Eq.~(3.14) in Ref.~\cite{2} in the limit
$P_B \gg M_B$.
This approximation deviates from the exact result by less than 11\% and 5\% for
$P_B=10$~GeV and 20~GeV, respectively.
In most of our applications, we have $P_B>20$~GeV.

\section{Results}
\label{sec:three}

\begin{figure*}
\vspace{-1.5cm}
\begin{tabular}{llll}
\includegraphics[width=4.0cm,bb=34 14 420 470]{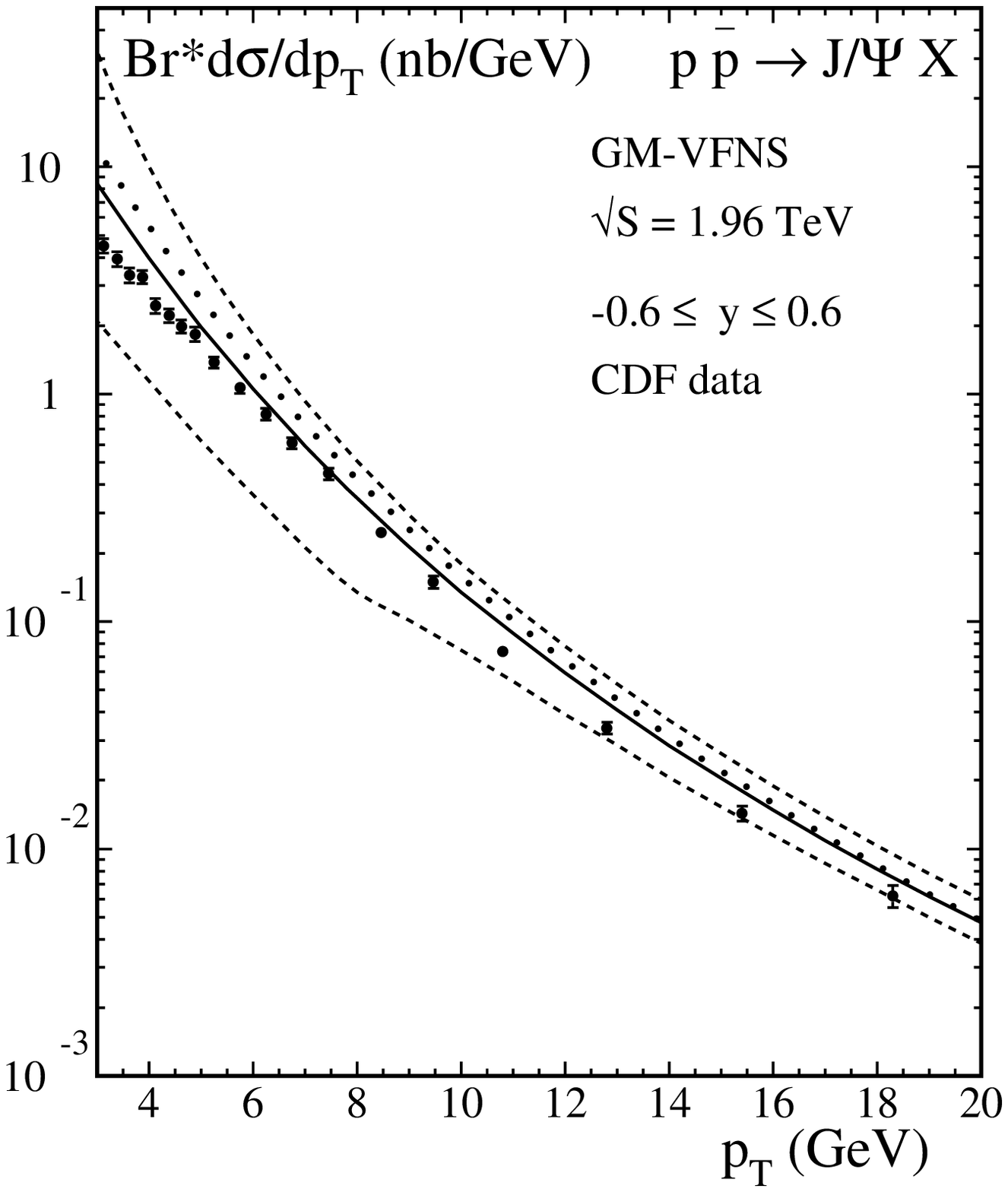} &
\includegraphics[width=4.0cm,bb=34 14 420 470]{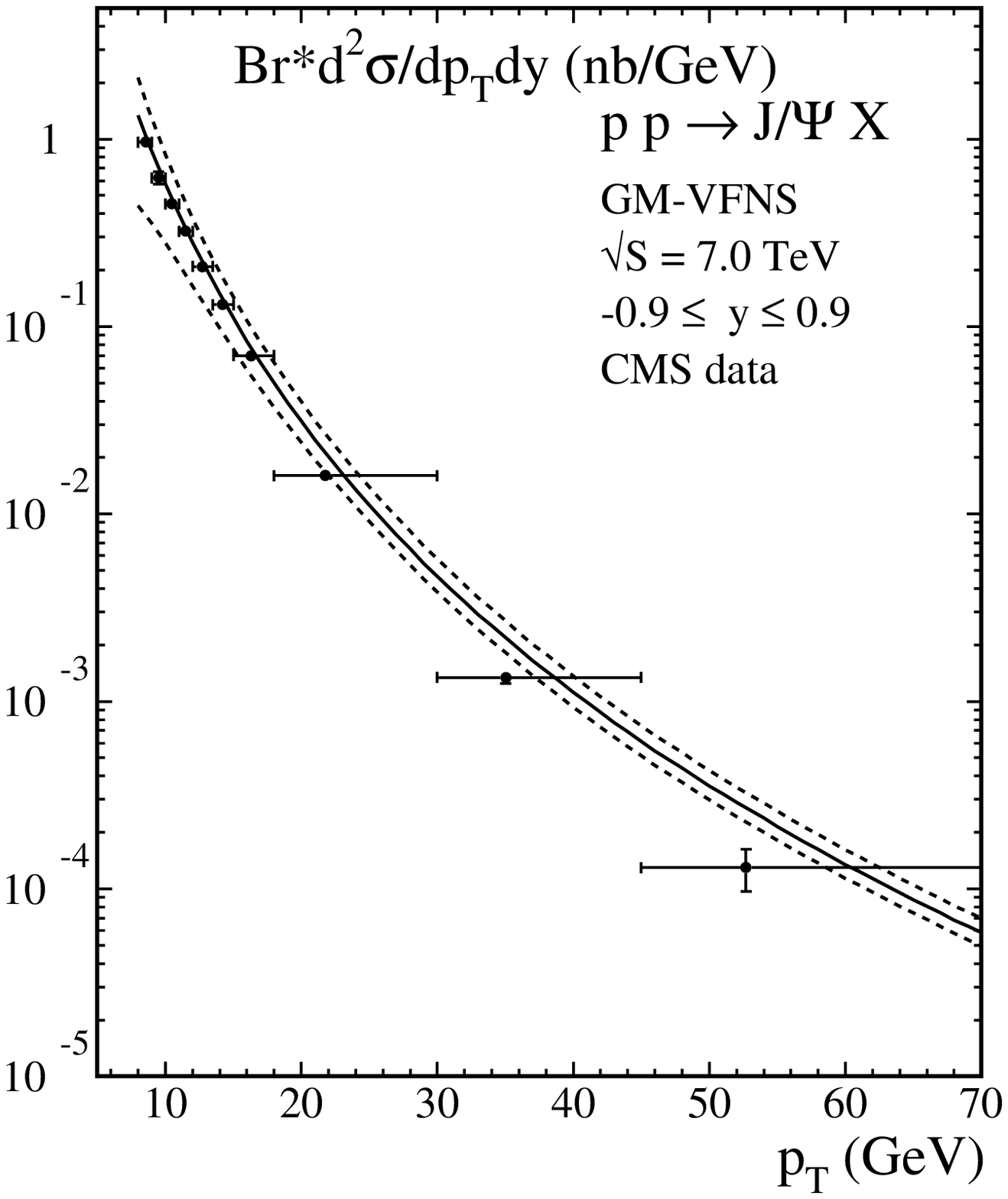} &
\includegraphics[width=4.0cm,bb=34 14 420 470]{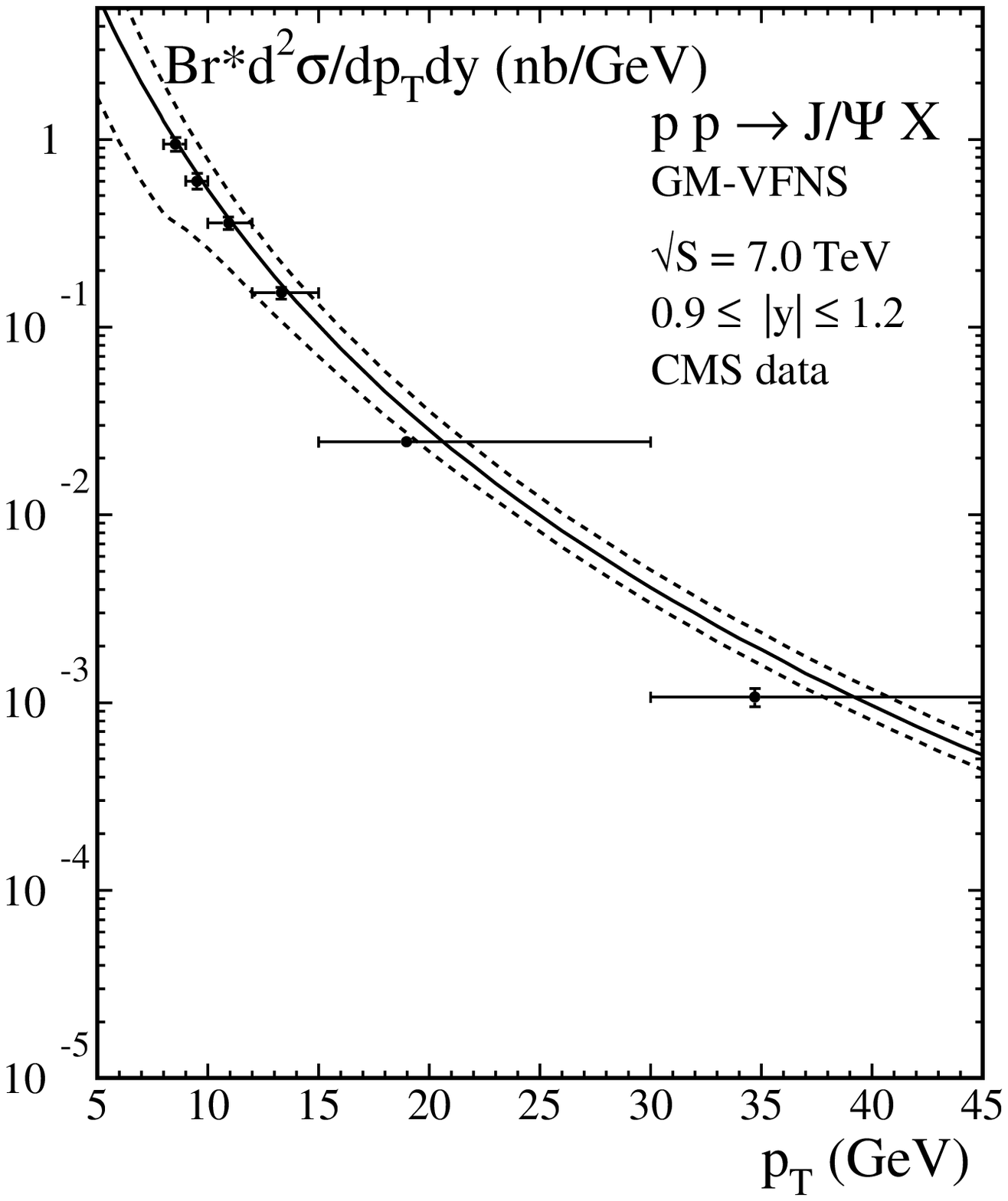} &
\includegraphics[width=4.0cm,bb=34 14 420 470]{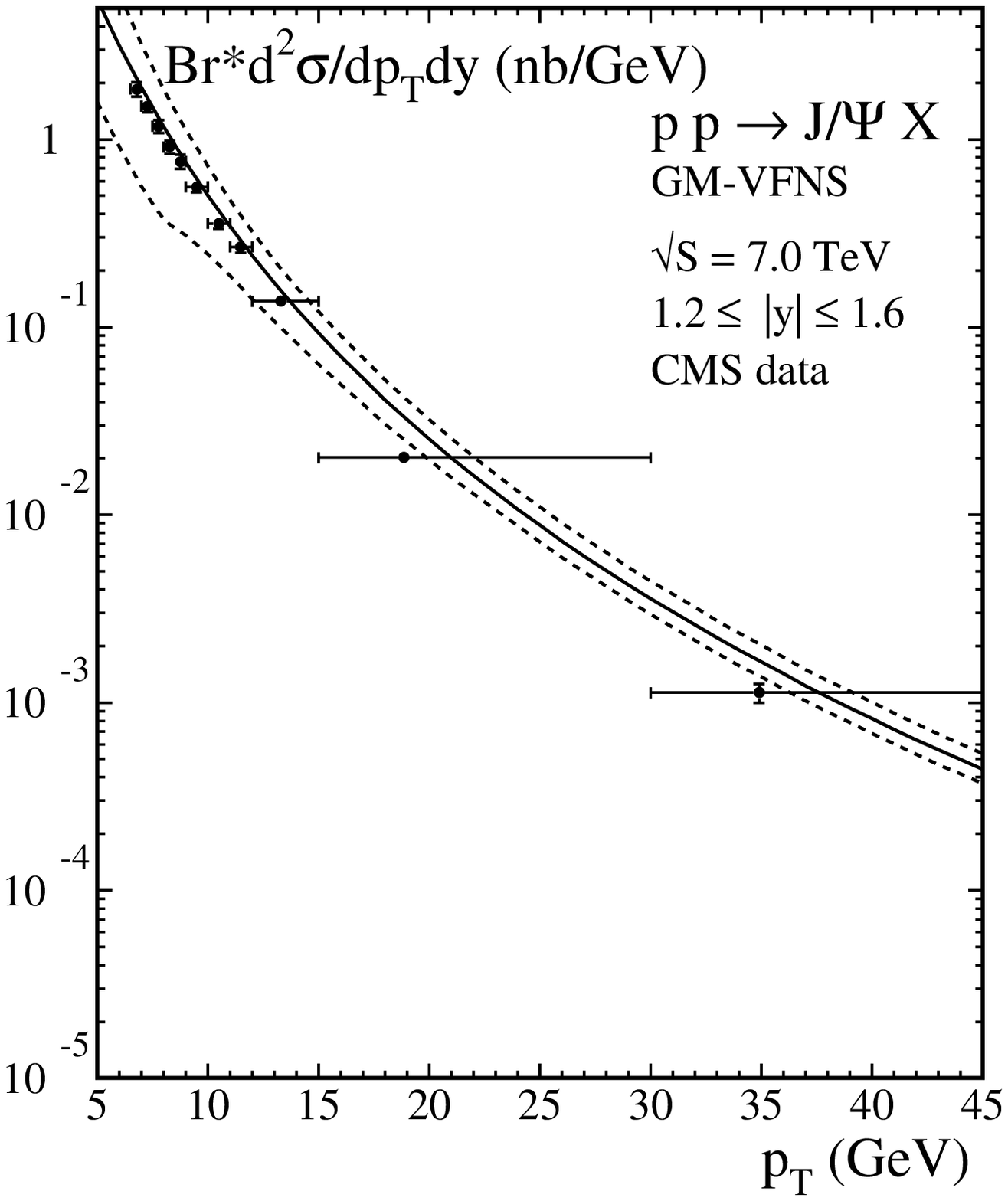}
\vspace{-0.7cm}
\\
(a) & (b) & (c) & (d)
\vspace{0.2cm}
\\
\includegraphics[width=4.0cm,bb=34 14 420 470]{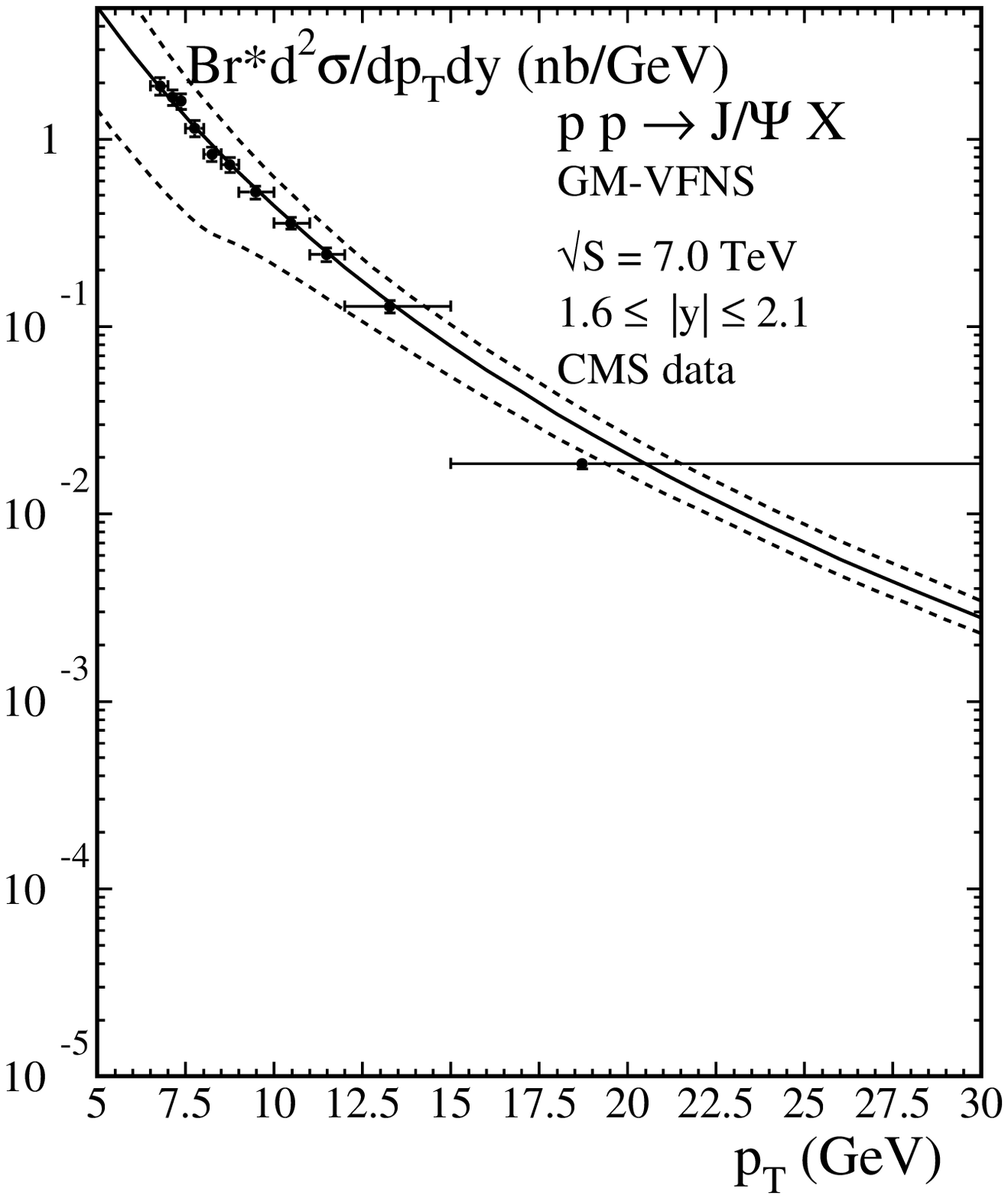} &
\includegraphics[width=4.0cm,bb=34 14 420 470]{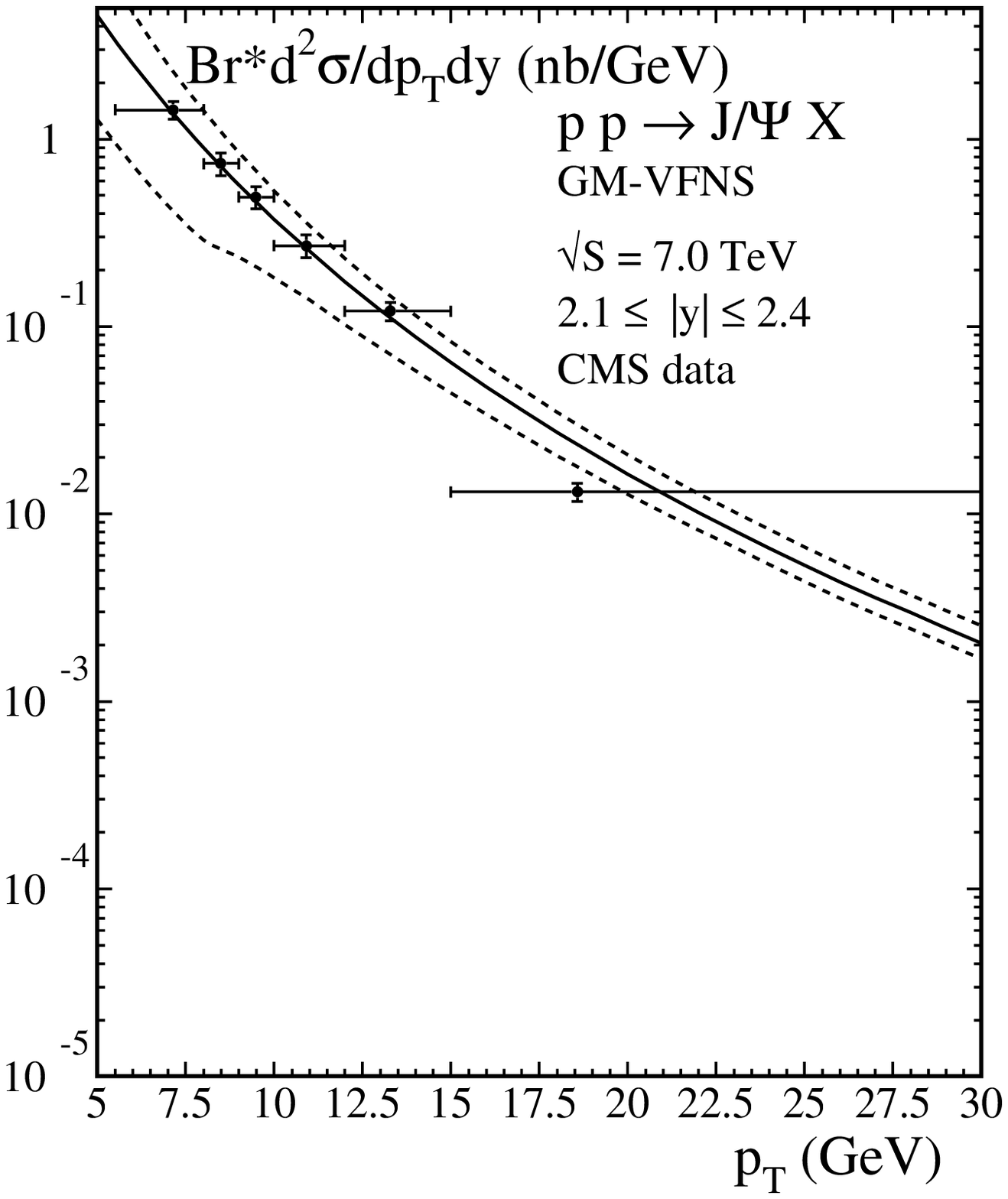} &
\includegraphics[width=4.0cm,bb=34 14 420 470]{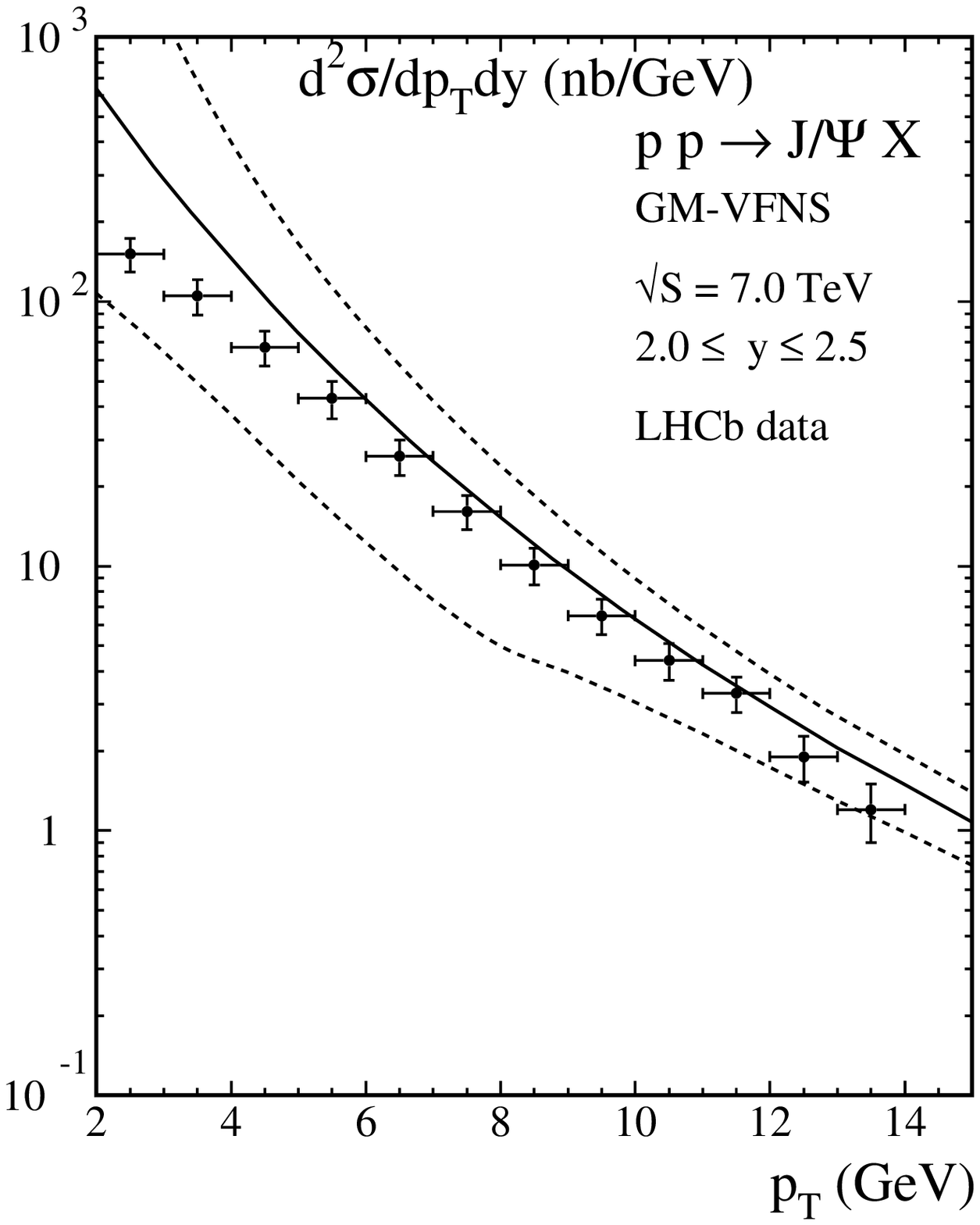} &
\includegraphics[width=4.0cm,bb=34 14 420 470]{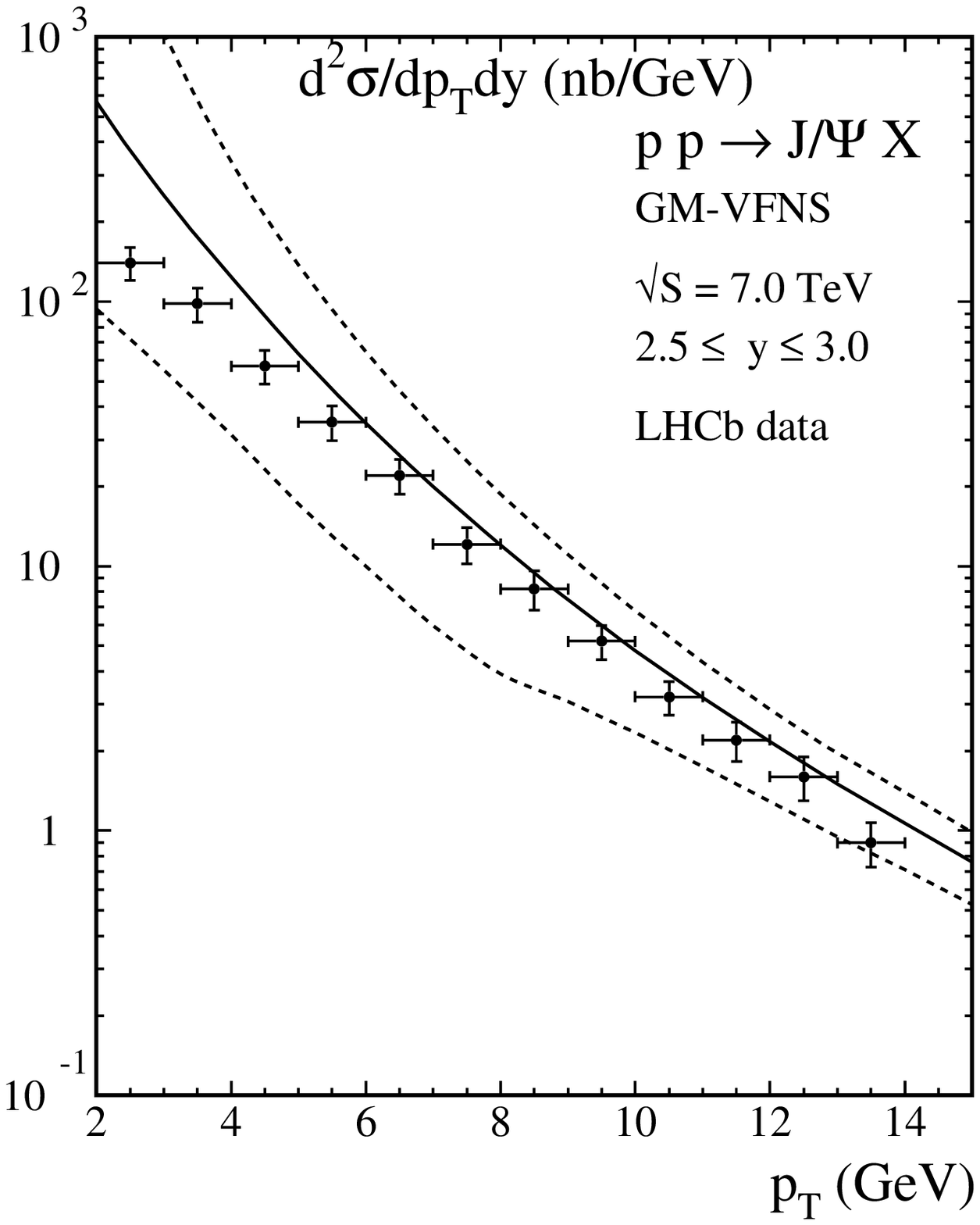}
\vspace{-0.7cm}
\\
(e) & (f) & (g) & (h)
\vspace{0.2cm}
\\
\includegraphics[width=4.0cm,bb=34 14 420 470]{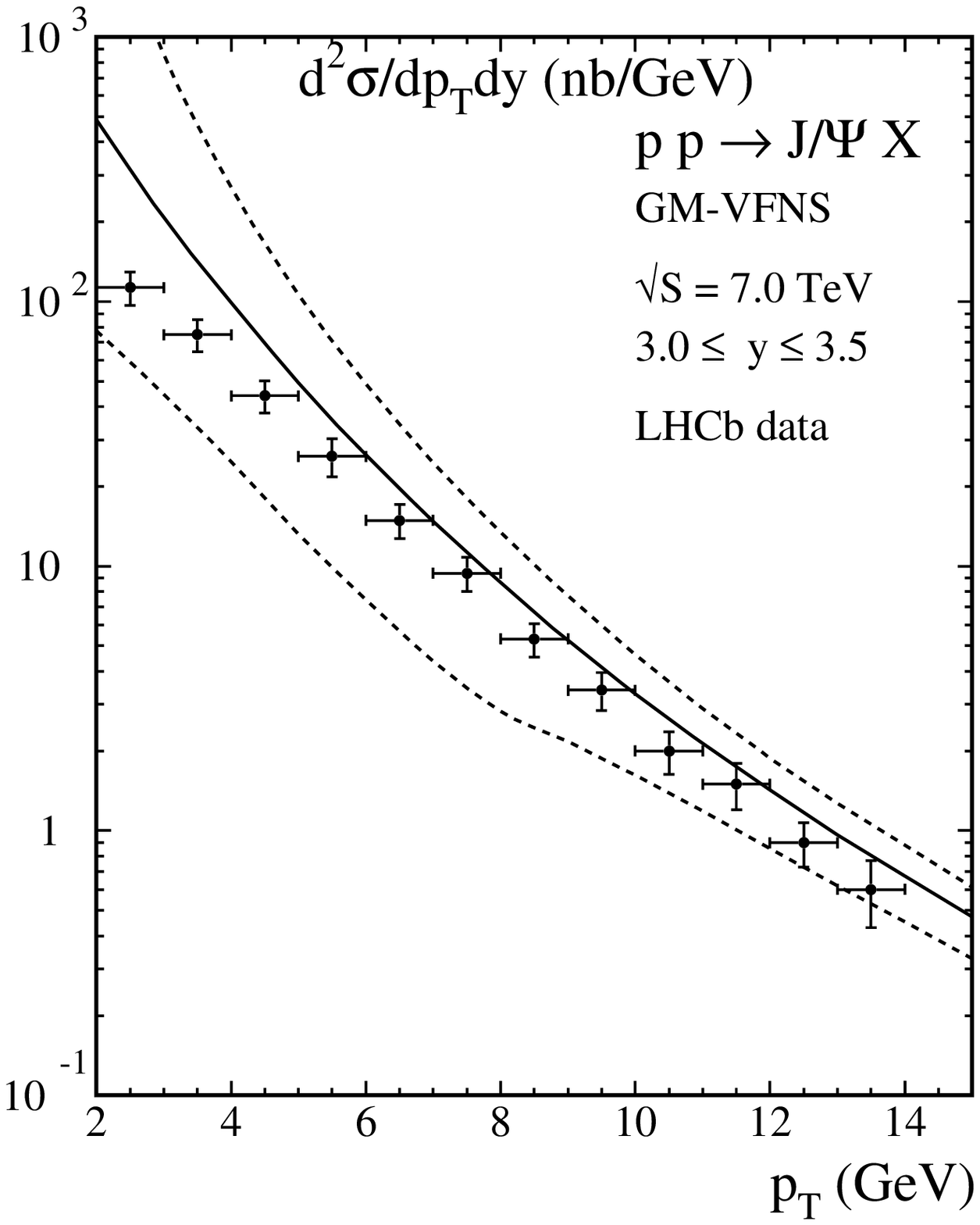} &
\includegraphics[width=4.0cm,bb=34 14 420 470]{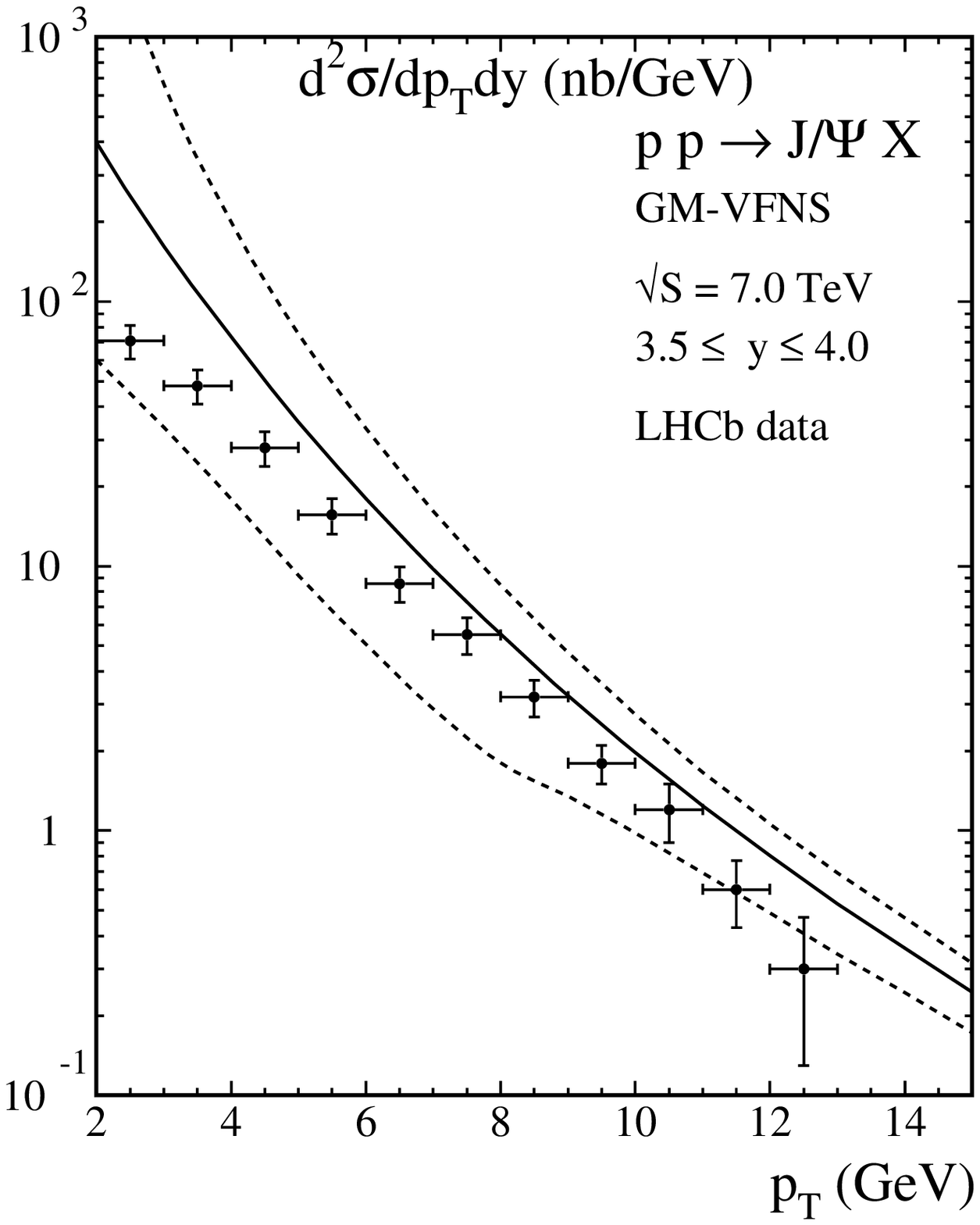} &
\includegraphics[width=4.0cm,bb=34 14 420 470]{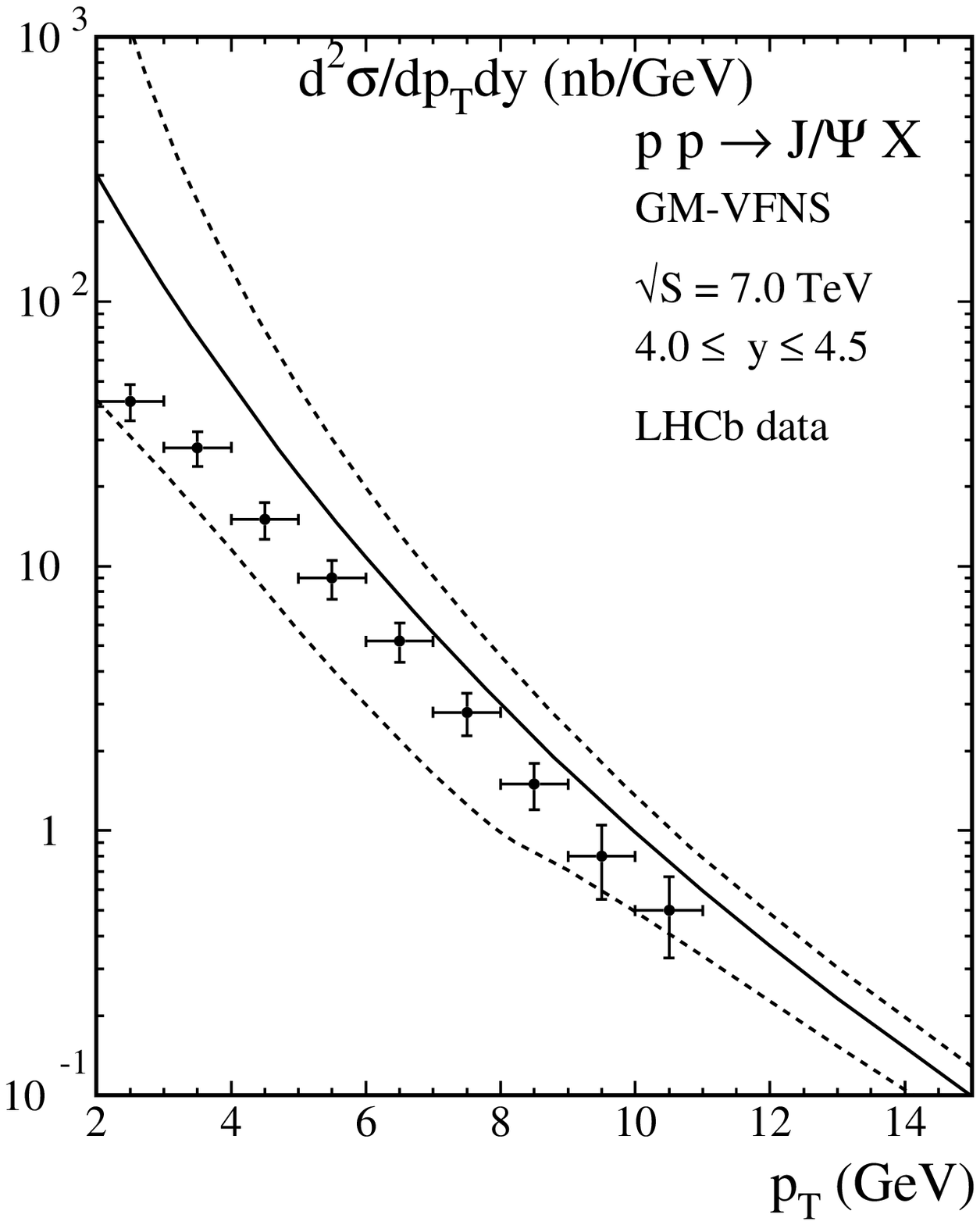} &
\includegraphics[width=4.0cm,bb=34 14 420 470]{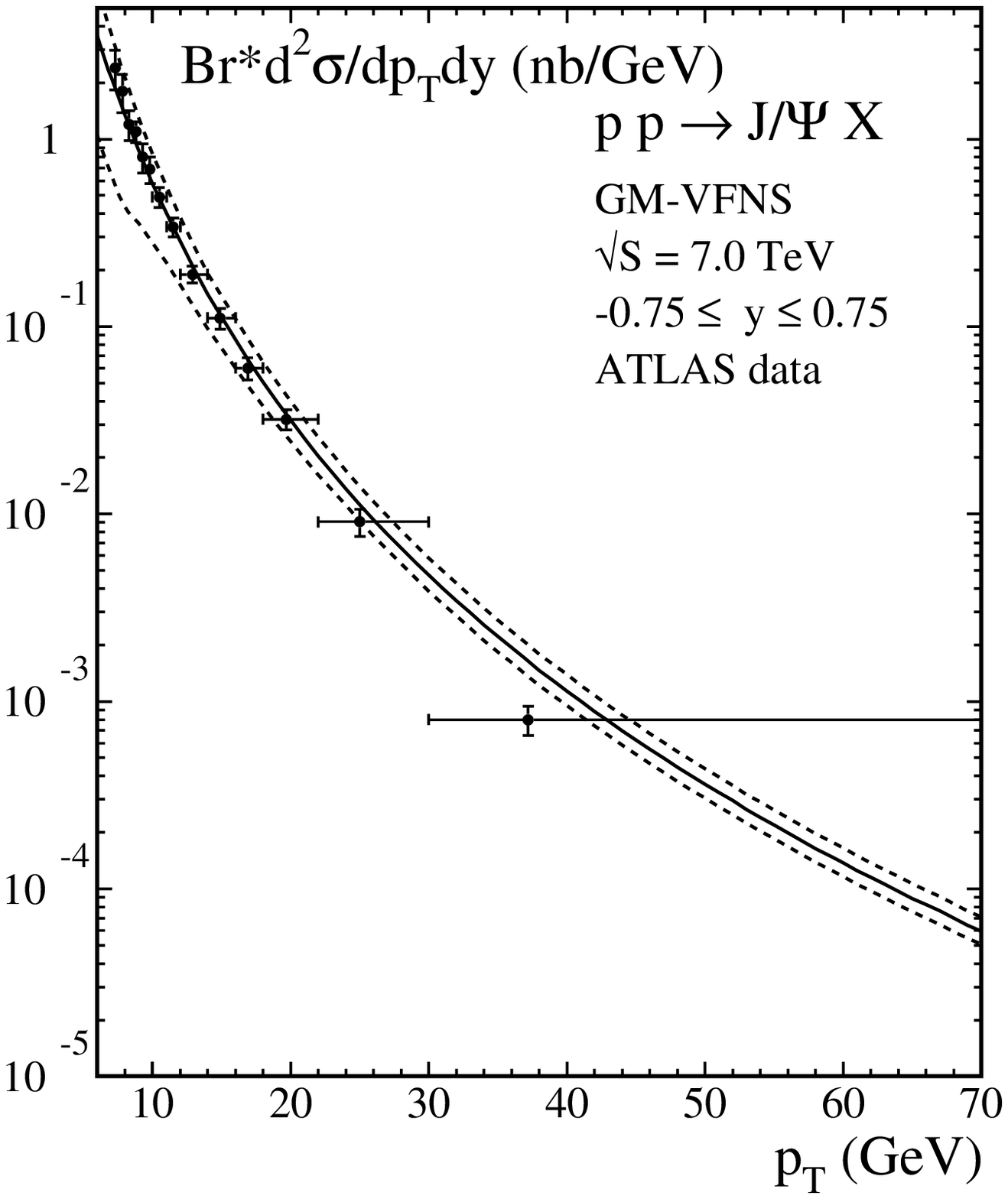}
\vspace{-0.7cm}
\\
(i) & (j) & (k) & (l)
\vspace{0.2cm}
\\
\includegraphics[width=4.0cm,bb=34 14 420 470]{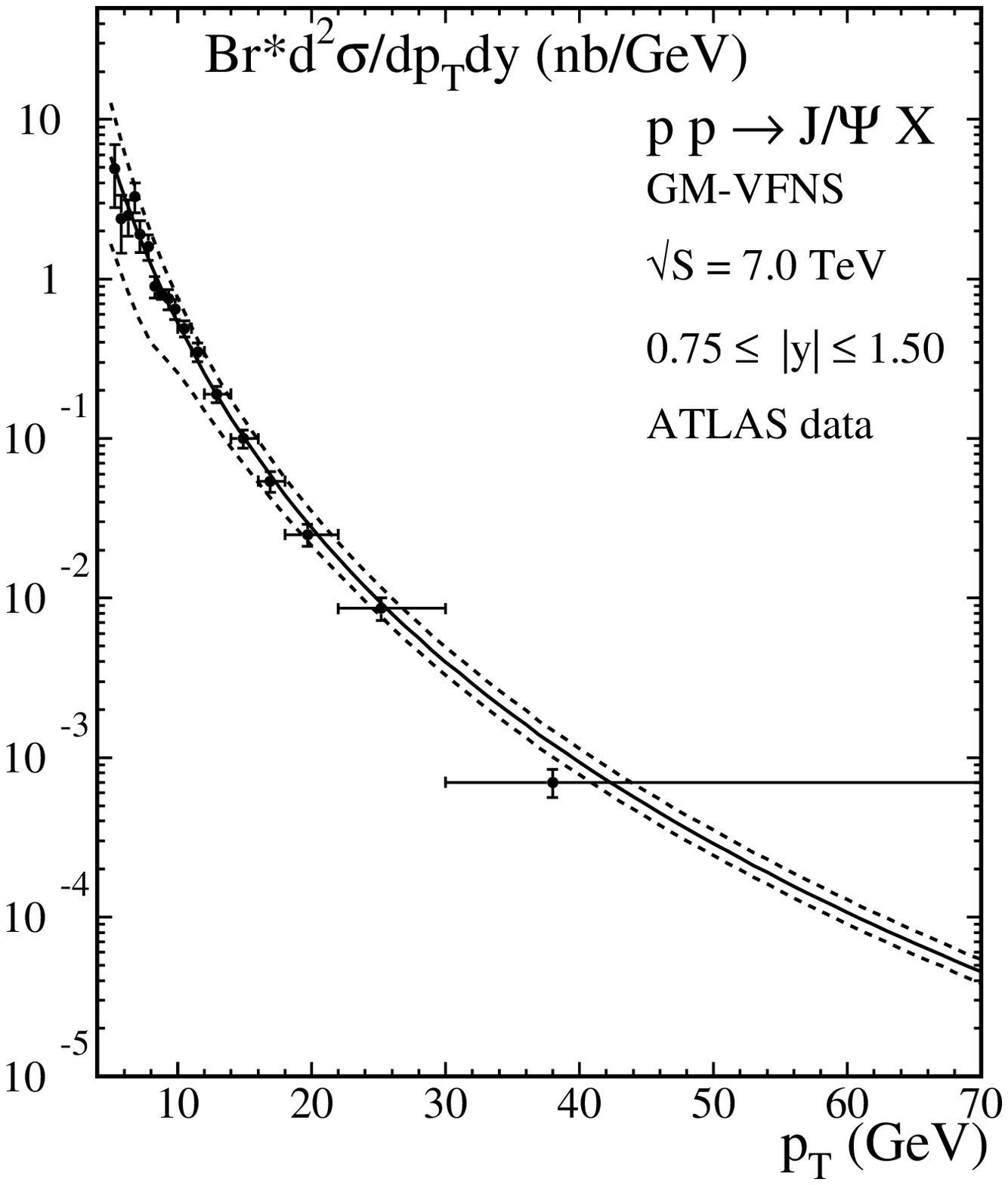} &
\includegraphics[width=4.0cm,bb=34 14 420 470]{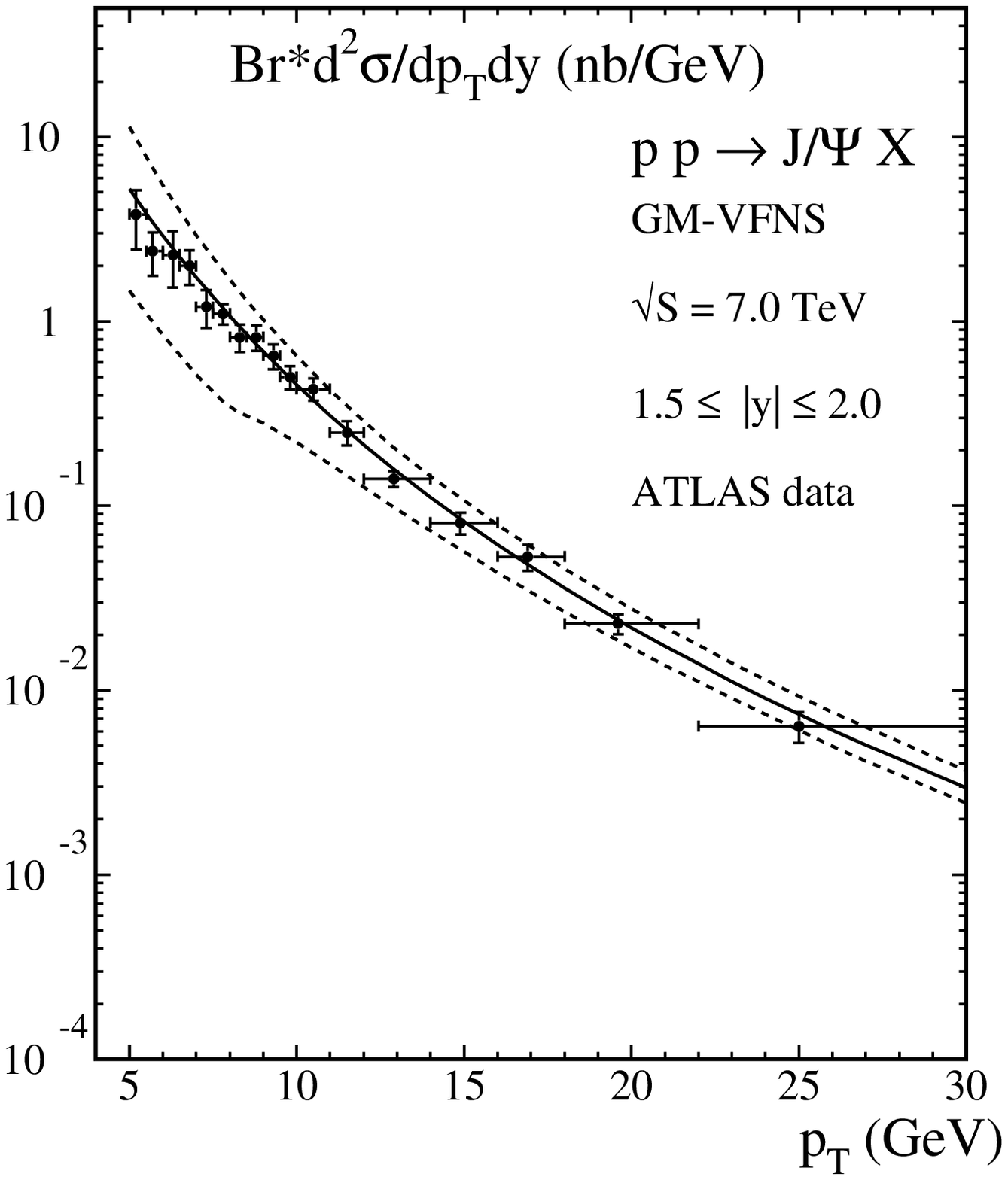} &
\includegraphics[width=4.0cm,bb=34 14 420 470]{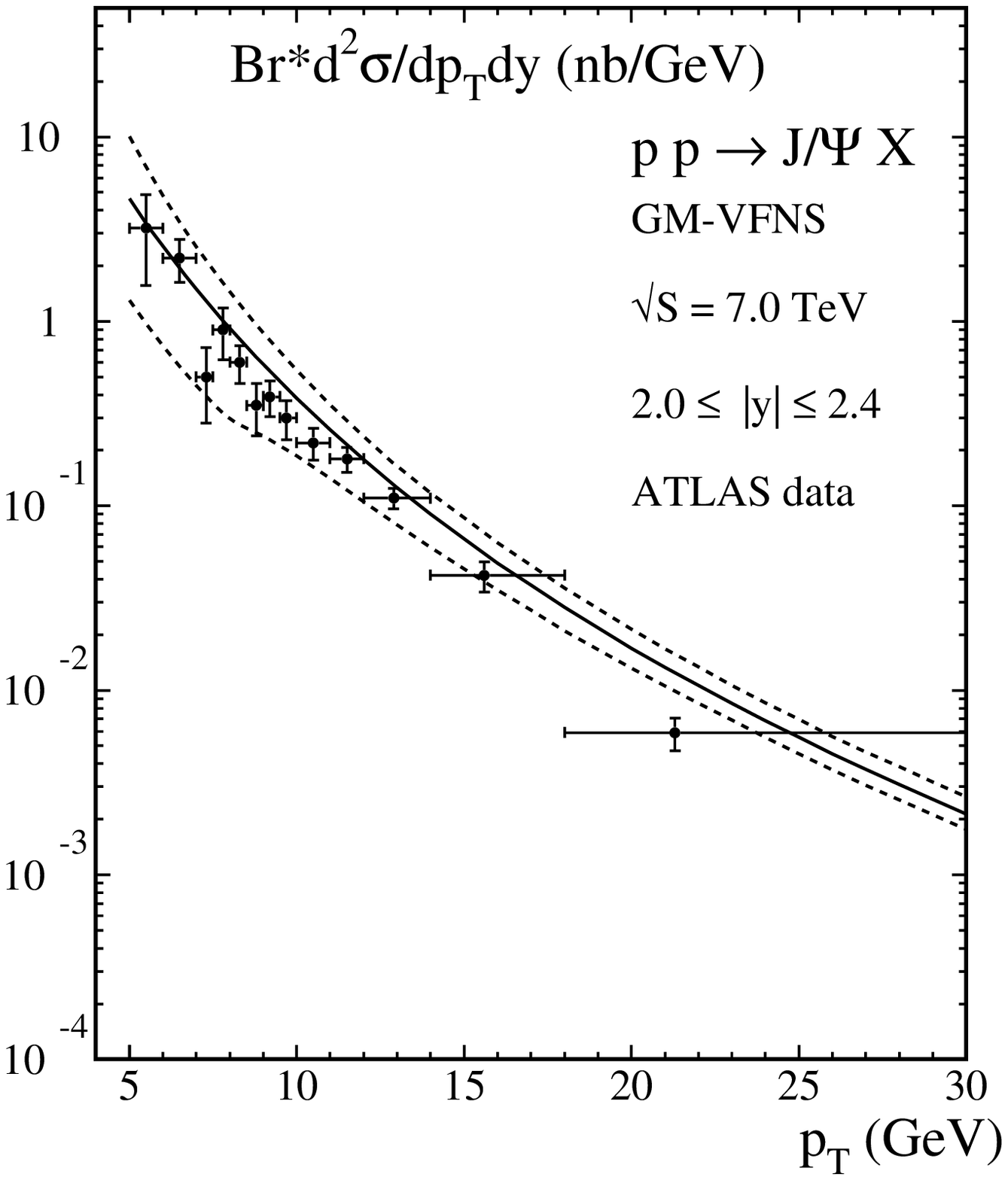} &
\includegraphics[width=4.0cm,bb=34 14 420 470]{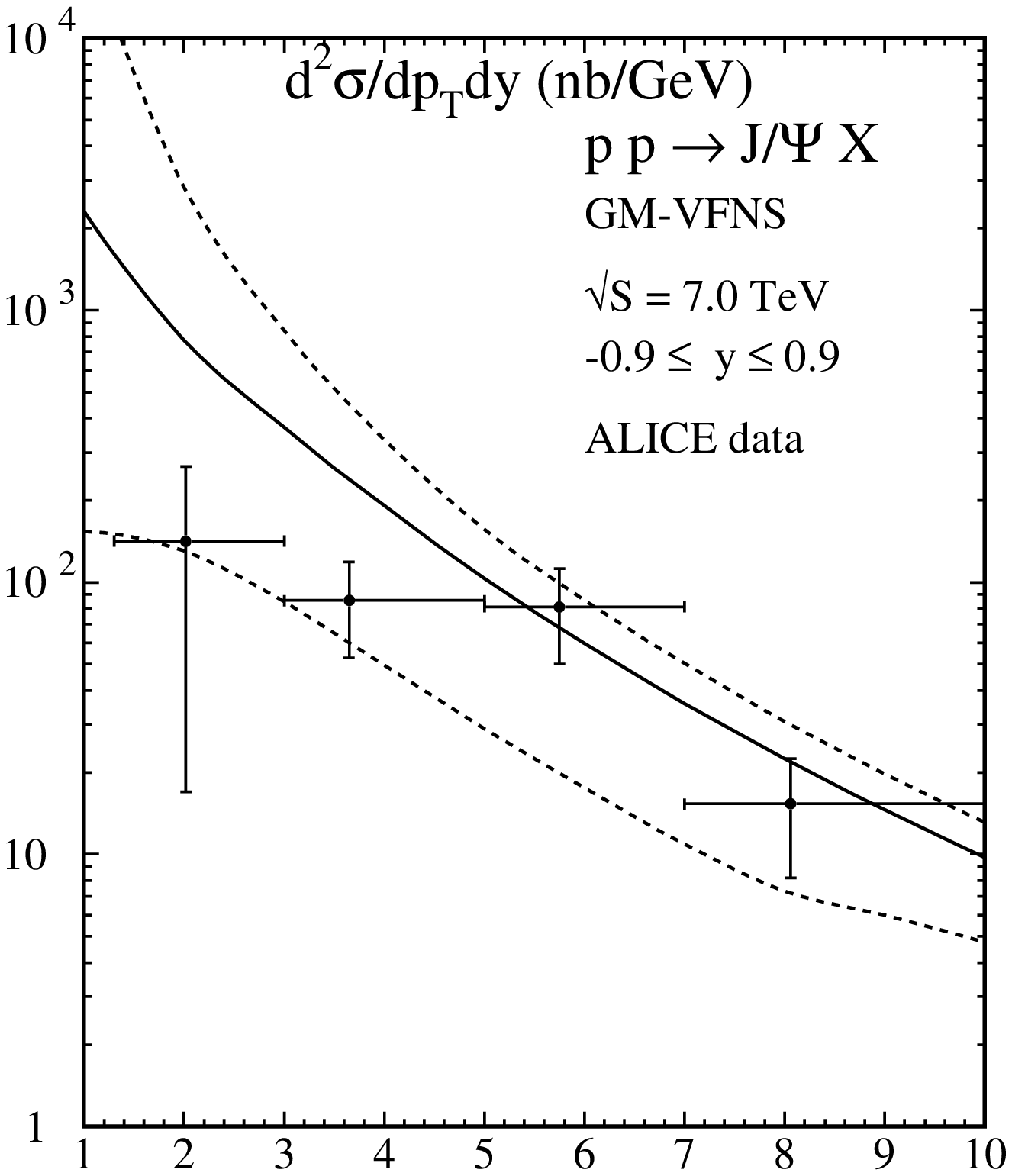}
\vspace{-0.7cm}
\\
(m) & (n) & (o) & (p)
\end{tabular}
\caption{\label{fig:one}The inclusive cross sections of nonprompt $J/\psi$
hadroproduction measured by CDF~II \cite{8} in $p\bar{p}$ collisions at
$\sqrt{s}=1.96$~TeV and by CMS \cite{9,10}, LHCb \cite{11}, ATLAS \cite{12},
and ALICE \cite{13} in $pp$ collisions at $\sqrt{s}=7$~TeV are compared with
NLO GM-VFNS predictions, whose default values and error bands are indicated by
the solid and dashed lines, respectively.
In frame~(a), the result obtained by replacing the total $B\to J/\psi+X$
three-momentum distribution in the default evaluation by a delta function in
$z$ peaking at $\langle z\rangle=0.6$ \cite{2} and normalized to the total
$B\to J/\psi+X$ branching fraction, 1.12\% \cite{2}, is represented by the
dotted line.}
\end{figure*}

\begin{figure*}
\begin{tabular}{lllll}
\includegraphics[width=3.15cm,bb=34 14 420 470]{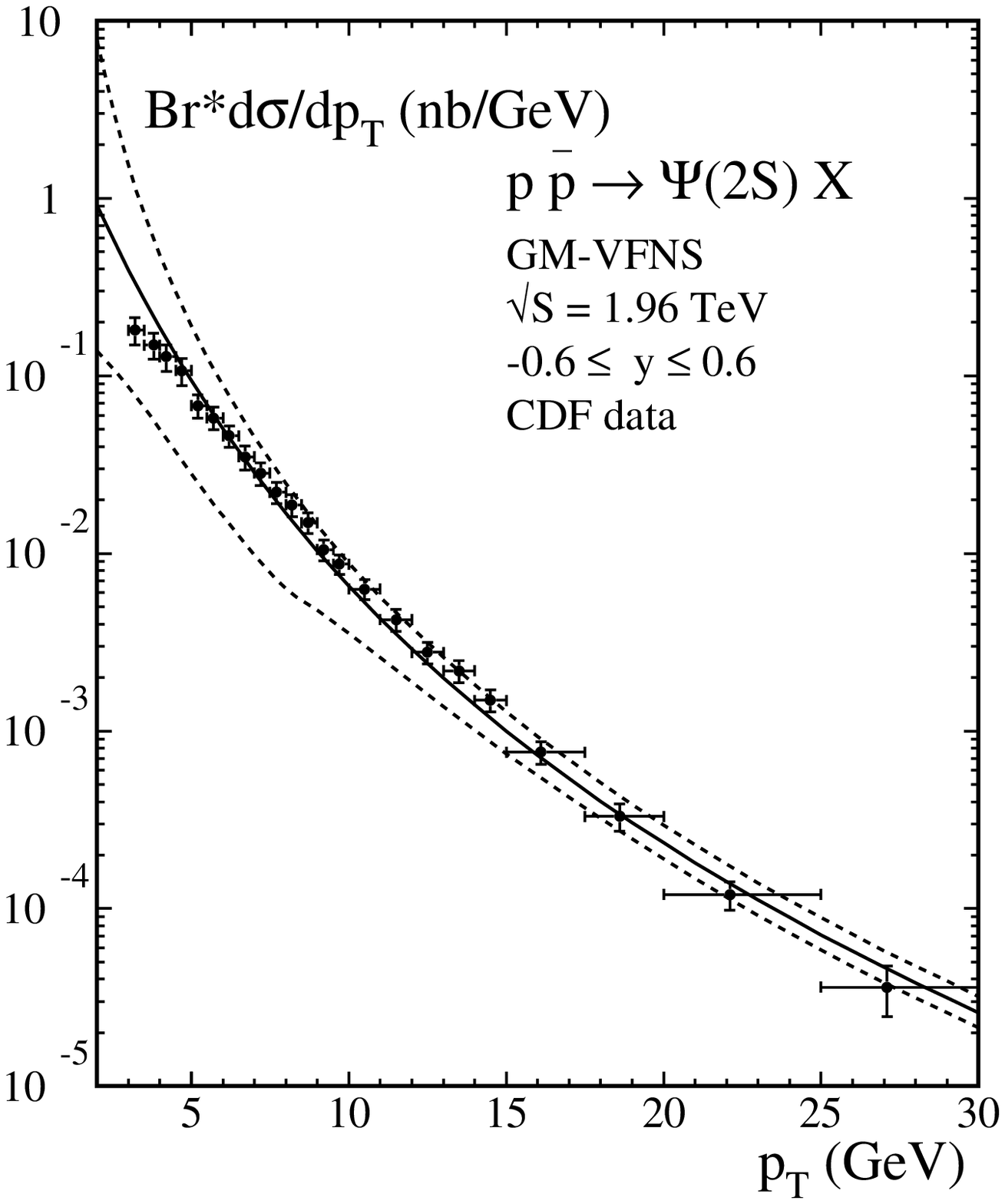} &
\includegraphics[width=3.15cm,bb=34 14 420 470]{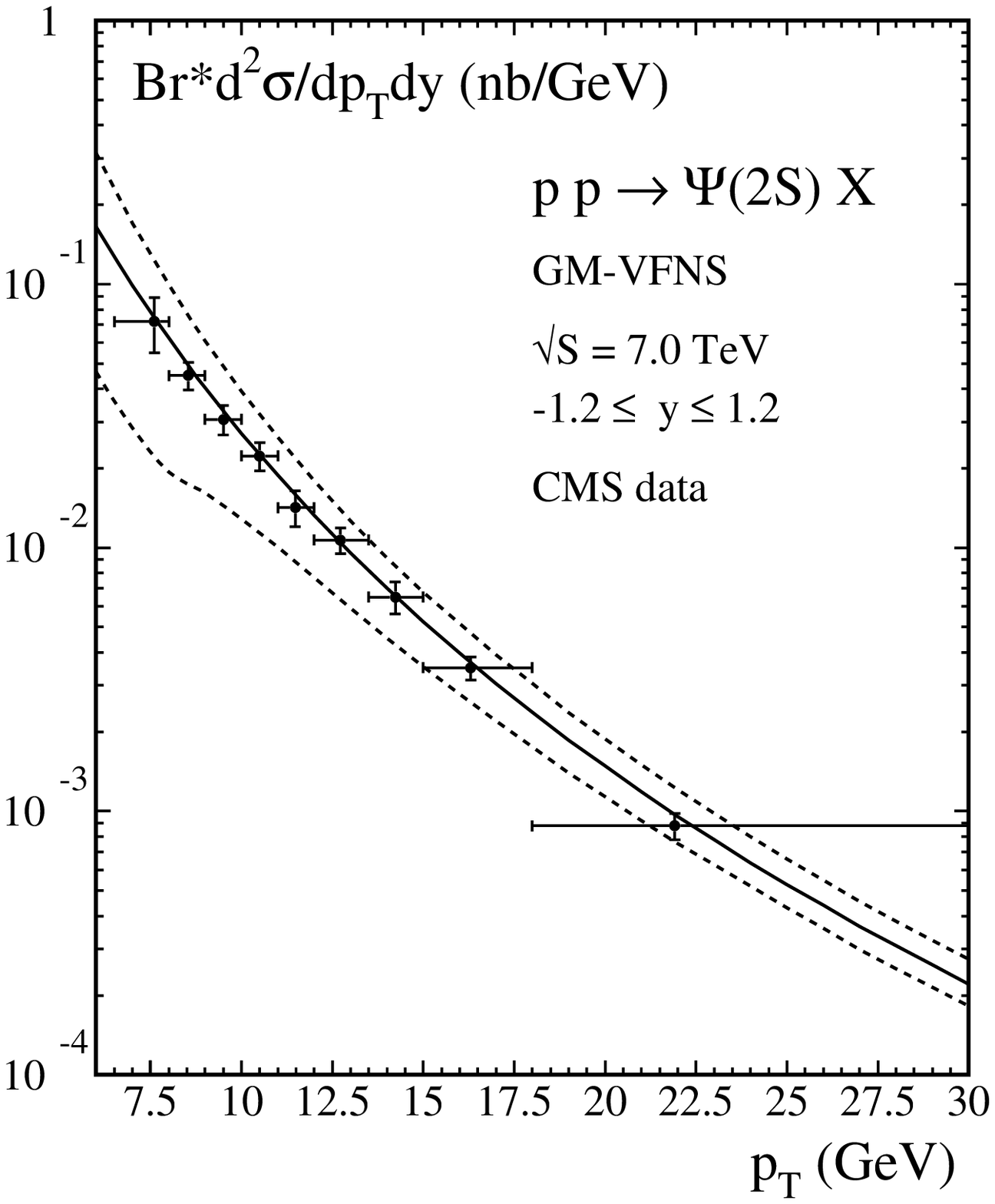} &
\includegraphics[width=3.15cm,bb=34 14 420 470]{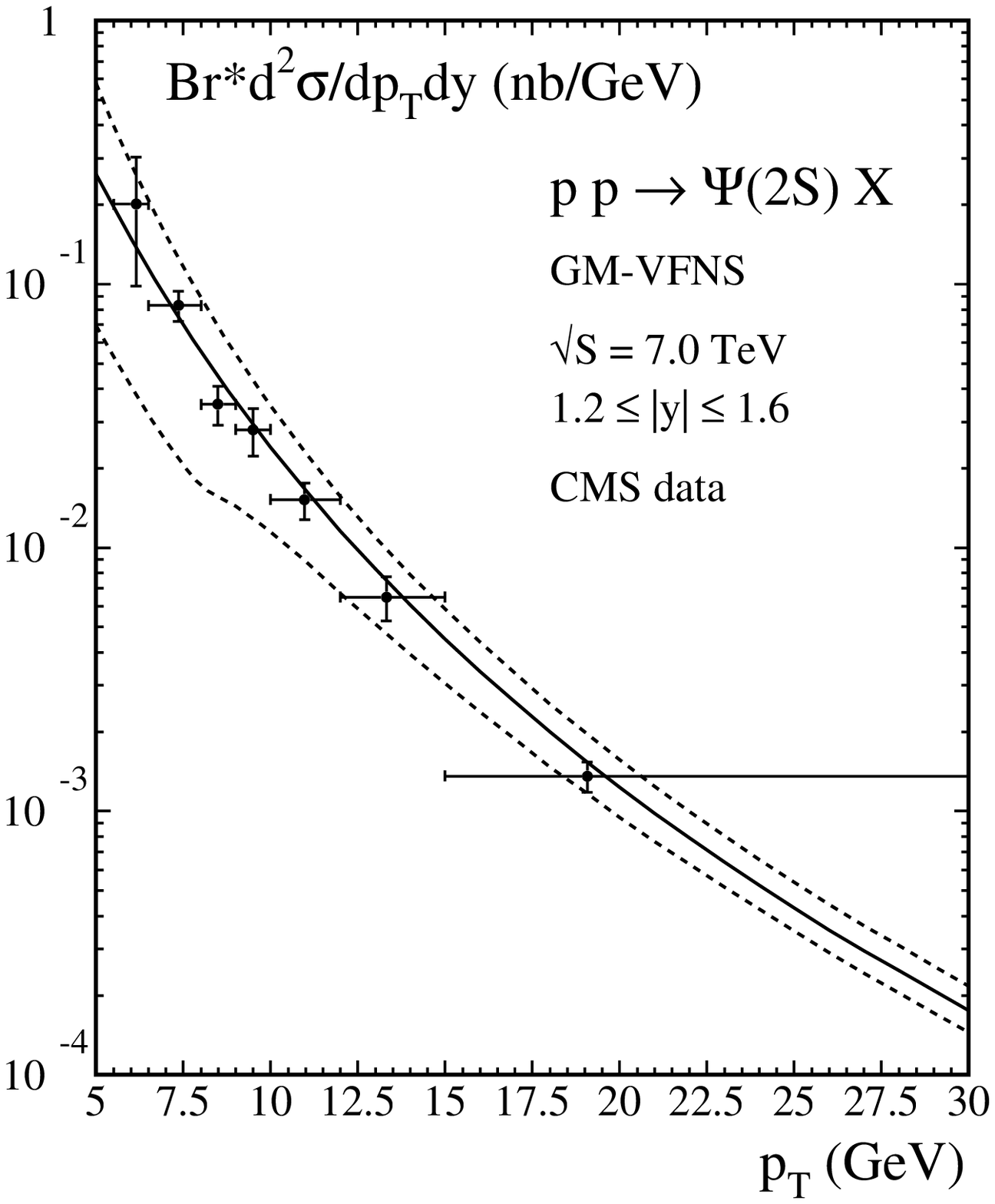} &
\includegraphics[width=3.15cm,bb=34 14 420 470]{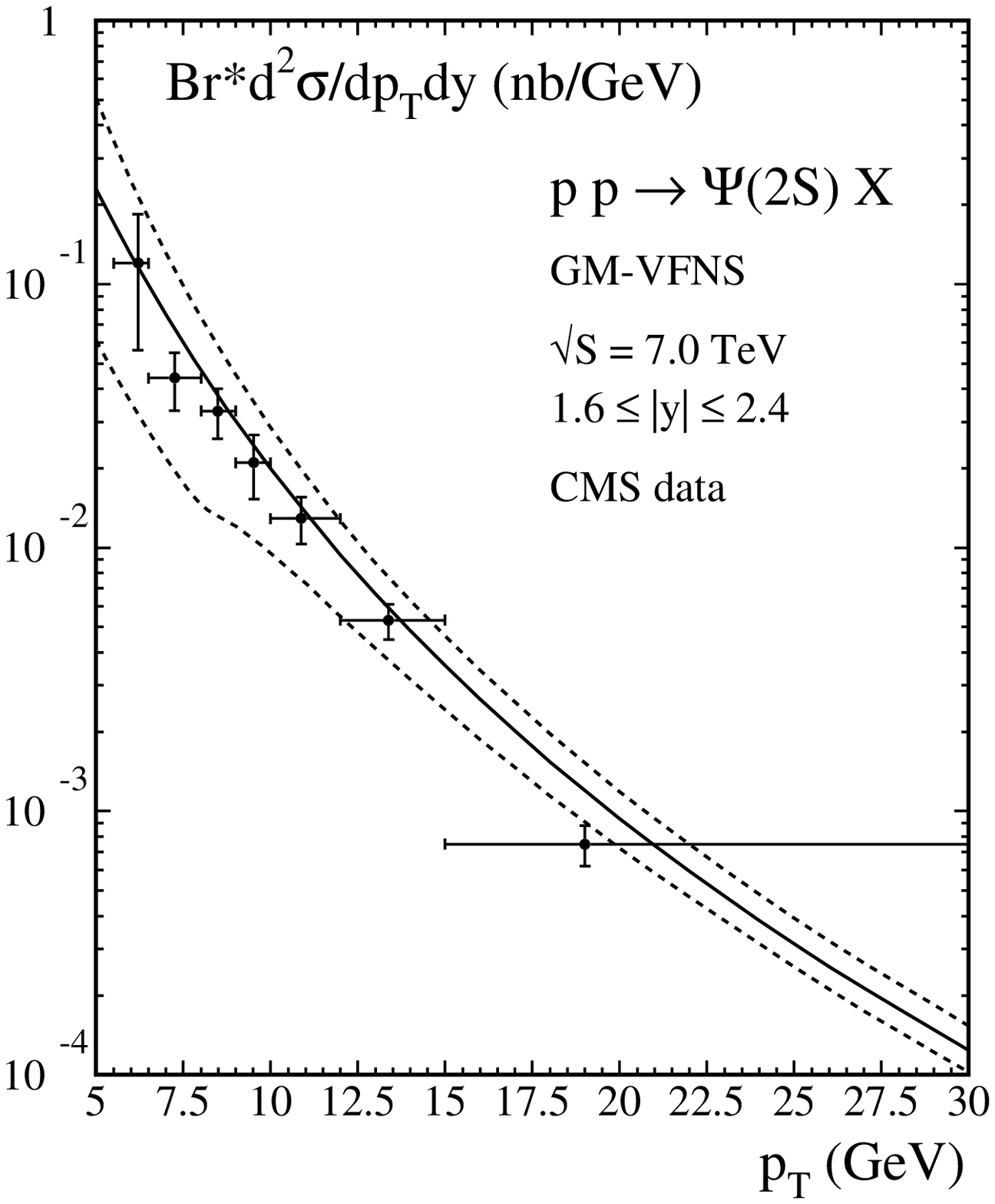} &
\includegraphics[width=3.15cm,bb=34 14 420 470]{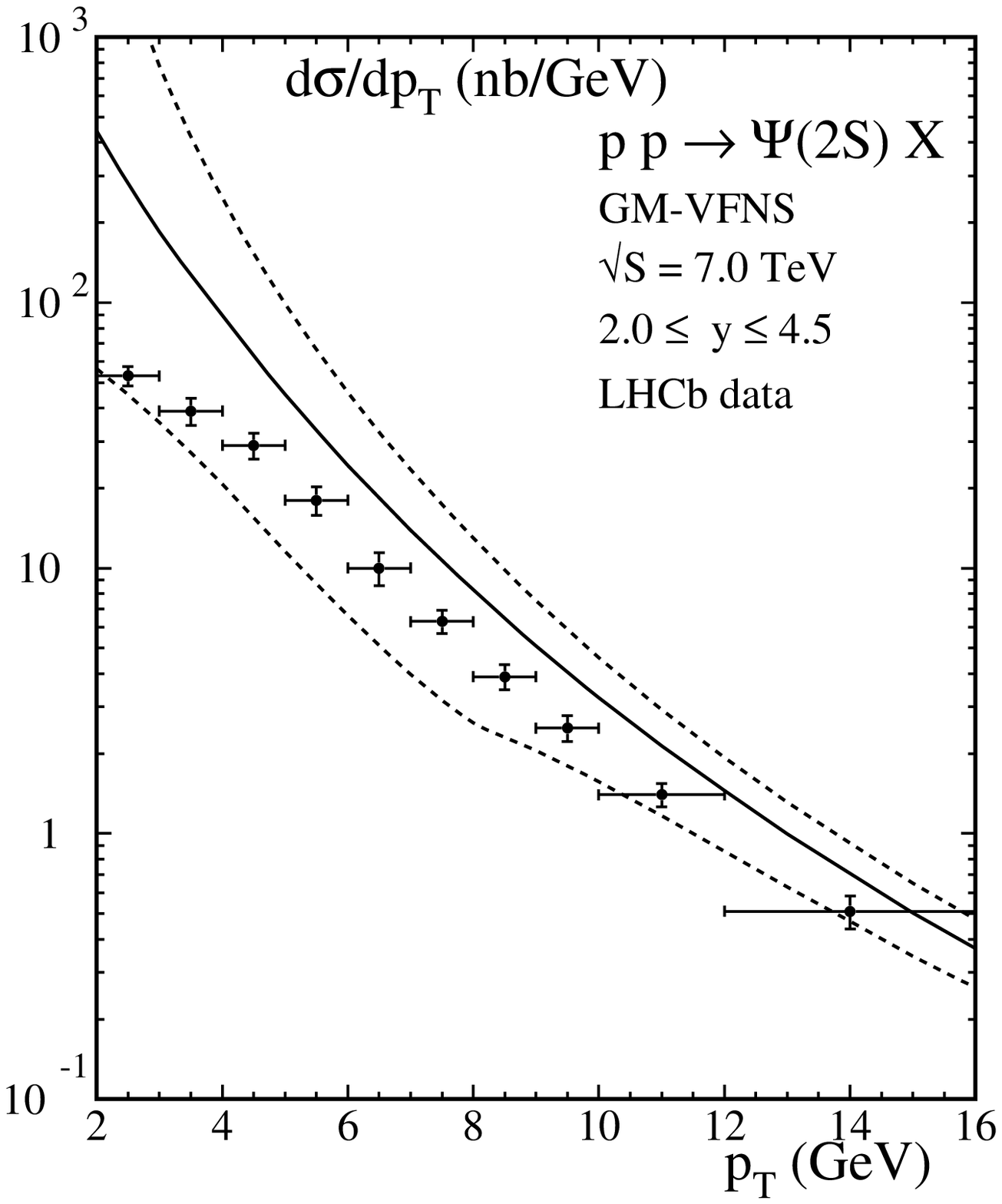}
\vspace{-0.6cm}
\\
(a) & (b) & (c) & (d) & (e)
\end{tabular}
\caption{\label{fig:two}The inclusive cross sections of nonprompt $\psi(2S)$
hadroproduction measured by CDF~II \cite{26} in $p\bar{p}$ collisions at
$\sqrt{s}=1.96$~TeV and by CMS \cite{10} and LHCb \cite{27} in $pp$ collisions
at $\sqrt{s}=7$~TeV are compared with NLO GM-VFNS predictions, whose default
values and error bands are indicated by the solid and dashed lines,
respectively.}
\end{figure*}

We are now in a position to present our numerical analysis.
In Figs.~\ref{fig:one} and \ref{fig:two}, we compare measurements of the
inclusive cross sections of nonprompt $J/\psi$ \cite{8,9,10,11,12,13} and
$\psi(2S)$ \cite{26,10,27} hadroproduction, respectively, with our NLO GM-VFNS
predictions evaluated as described in Sec.~\ref{sec:two}. 
The experimental data come as the cross section distributions
$d\sigma/dp_T$ integrated over $2.0<y<4.5$ \cite{27},
$B\times d\sigma/dp_T$ integrated over $|y|<0.6$ \cite{8,26},
$d^2\sigma/(dp_Tdy)$ \cite{11,13}, and
$B\times d^2\sigma/(dp_Tdy)$ \cite{9,12},
where $B$ stands for the branching fractions of the decays $J/\psi\to\mu^+\mu^-$
and $\psi(2S)\to\mu^+\mu^-$, for which we adopt the values $B=5.93\%$ and
0.77\%, respectively, from Ref.~\cite{24}.
Besides the default predictions with $\xi_F=\xi_R=1$, we also present error
bands encompassed between the minimum and maximum values obtained by the
variations of $\xi_F$ and $\xi_R$ as explained in Sec.~\ref{sec:two}.
The slight changes of slope in the lower bounds at about $p_T=8$~GeV reflect
the fact that the partonic subprocesses initiated by a $b$ quark are turned off
by the $b$-quark PDF as the threshold at $\mu_F=m_b$ is reached.

We now take a closer look at Fig.~\ref{fig:one}.
From Fig.~\ref{fig:one}(a), we observe that the CDF~II data points \cite{8} are
all contained within the theoretical-error band, exhibiting a slight tendency
to undershoot the default prediction at small and large $p_T$ values.
We do not consider data available in the range $1.25<p_T<3.0$~GeV \cite{8},
where our theoretical predictions are less reliable.
In order to illustrate the importance of a realistic description of the
total $B\to J/\psi+X$ three-momentum distribution, we repeat the default
evaluation after replacing in Eq.~(\ref{eq:ff})
\begin{equation}
\frac{1}{\Gamma_B}\,\frac{d\Gamma(z,P_B)}{dz}=B\,\delta(z-\langle z\rangle),
\end{equation}
where $B=1.12\%$ \cite{2} is the total $B\to J/\psi+X$ branching fraction and
$\langle z\rangle=0.6$ is the average value of $z$ read off from Fig.~3 in
Ref.~\cite{2}.
The result, which may be simply evaluated as
\begin{equation}
\frac{d\sigma}{dp_T}(p\bar p\to J/\psi+X)=\frac{B}{\langle z\rangle}\,
\frac{d\sigma}{d(p_T/\langle z\rangle)}(p\bar p\to B+X),
\end{equation}
overshoots the default prediction by as much as 40\% at $p_T=3$~GeV, but
smoothly merges with the latter as the value of $p_T$ approaches 20~GeV.
Similarly, switching from the GM-VFNS \cite{16,14,15} to the ZM-VFNS
\cite{4} has an appreciable effect only at small values of $p_T$, provided the
$b\to B$ FF \cite{14} is maintained, as may be inferred from Figs.~7 and 8 in
Ref.~\cite{14}.
The CMS data \cite{9,10} shown in Figs.~\ref{fig:one}(b)--(f) are sampled in
the five $y$ bins $|y|<0.9$, $0.9<|y|<1.2$, $1.2<|y|<1.6$, $1.6<|y|<2.1$, and
$2.1<|y|<2.4$, respectively, and cover different $p_T$ ranges.
The measurement in the most central rapidity bin reaches out through
$p_T=70$~GeV.
The experimental errors shown are obtained, for simplicity, by summing
quadratically the statistical, systematic, and luminosity-related errors, with
the understanding that this procedure is likely to overestimate the uncertainty
in the lineshape of the $p_T$ distribution because the luminosity-related
errors are correlated among the individual data points and mainly affect the
overall normalization.
The agreement between experiment and theory is rather satisfactory, except
for the largest-$p_T$ bins, where the measurements including their errors tend
to lie underneath the theory bands.
The LHCb data \cite{11} displayed in Figs.~\ref{fig:one}(g)--(k) refer to five
$y$ bins of equal widths in the range $2.0< y< 4.5$ covering different $p_T$
ranges, the widest being 2.0~GeV${}<p_T<14.0$~GeV. 
With one exception, all the central data points fall inside the theory bands.
The data points tend to undershoot the default predictions, the more so at
small $p_T$ values.
The ATLAS data \cite{12} included in Figs.~\ref{fig:one}(l)--(o) are grouped in
the four $y$ bins $|y|<0.75$, $0.75<|y|<1.5$, $1.5<|y|<2.0$, and $2.0<|y|<2.4$,
respectively, and cover $p_T$ values as large as 70~GeV.
They agree very well with our NLO GM-VFNS predictions, being gathered within
the theory bands, with the exception of the data points of largest $p_T$ in
each of Figs.~\ref{fig:one}(l), (m), and (o), which are slightly below.
In fact, most of the data points even agree with our default predictions within
the experimental errors.
Very recently, the ALICE Collaboration reported their measurement of prompt and
nonprompt $J/\psi$ hadroproduction in Ref.~\cite{13}. 
There are four ALICE data points, in the kinematic range $p_T>1.3$~GeV and
$|y|<0.9$, which may be extracted from Ref.~\cite{13} by multiplying the
respective results for the inclusive cross section of prompt plus nonprompt
$J/\psi$ hadroproduction and the fraction of $J/\psi$ mesons from $b$-hadron
decays, appropriately combining the experimental errors. 
All the four data points agree with our NLO GM-VFNS predictions within the
theoretical uncertainties as may be seen in Fig.~\ref{fig:one}(p).

We now move on to Fig.~\ref{fig:two}.
While nonprompt $J/\psi$ production is also possible via the feed-down from
heavier charmonia, nonprompt $\psi(2S)$ production proceeds only directly.
The CDF~II data \cite{26}, the CMS data \cite{10} in the $y$ bins $|y|<1.2$,
$1.2<|y|<1.6$, and $1.6<|y|<2.4$, and the LHCb data \cite{27} are compared
with our NLO GM-VFNS predictions in Figs.~\ref{fig:two}(a)--(e), respectively.
The CDF~II and CMS measurements, in the central regions of the detectors, reach
out to $p_T=30$~GeV, while the LHCb one, in the forward region, stops at
$p_T=16$~GeV. 
We conclude from Fig.~\ref{fig:two} that all the experimental data points agree
with our NLO GM-VFNS predictions within the theoretical uncertainties.
With a few exceptions, all the CDF~II and CMS data points agree with our
default predictions within the experimental errors, while the LHCb data points
consistently undershoot our default predictions.

\section{Conclusions}
\label{sec:four}

Motivated by recent measurements at the Tevatron \cite{8,26} and the LHC
\cite{9,10,11,27,12,13}, we improved and updated our previous analysis of the
inclusive cross sections of nonprompt $J/\psi$ and $\psi(2S)$ hadroproduction
\cite{2} by adopting the GM-VFNS \cite{16,14,15} and refreshing our inputs as
described in Sec.~\ref{sec:two}.
In Sec.~\ref{sec:three}, the transverse-momentum distributions measured by the
CDF~II \cite{8,26}, CMS \cite{9,10}, LHCb \cite{11,27}, ATLAS \cite{12}, and
ALICE \cite{13} Collaborations were found to be very well described by our
upgraded NLO predictions, as for both absolute normalization and lineshape.
This constitutes a nontrivial test of the GM-VFNS over wide $\sqrt{s}$, $p_T$,
and $y$ ranges.

\section*{Acknowledgements}

This work was supported by the German Federal Ministry for Education and
Research BMBF through Grant No.\ 05H12GUE and by the German Research Foundation
DFG through Grant No.\ KN~365/5-3.

\end{document}